\documentclass[preprint2]{aastex}

\def\be{\begin{equation}}
\def\ee{\end{equation}}

\def\etal{ et al. }
\def\hi{{\ion{H}{1}}}

\newcommand{\kms}{\ensuremath{\rm{km\,s^{-1}}}}

\newcommand{\dg}{\ensuremath{^{\circ}}}
\newcommand{\jykms}{\ensuremath{\mathrm{Jy} \,\kms}}
\newcommand{\cm}{\ensuremath{\mathrm{cm}^{-2}}}
\def\msun{\ensuremath{M_\odot}}

\def\sn{\ensuremath{S/N}}
\def\nh{\ensuremath{N_{HI}}}
\def\nhp{\ensuremath{\nh^{\prime}}}
\def\chan{\ensuremath{\delta V}}
\def\chanp{\ensuremath{\chan^{\prime}}}
\def\sm{\ensuremath{\sigma_{m}}}
\def\smp{\ensuremath{\sm^{\prime}}}
\def\mhi{\ensuremath{M_{HI}}}
\def\mhip{\ensuremath{\mhi^{\prime}}}
\def\wf{\ensuremath{W_{50}}}
\def\wfp{\ensuremath{\wf^{\prime}}}
\def\fc{\ensuremath{F_c}}
\def\fcp{\ensuremath{\fc^{\prime}}}
\def\cz{\ensuremath{cz}}
\def\czsun{\ensuremath{\cz_{\odot}}}
\def\czsunp{\ensuremath{\czsun^{\prime}}}
\def\mdyn{\ensuremath{M_{dyn}}}
\def\mdynp{\ensuremath{\mdyn^{\prime}}}
\def\ahi{\ensuremath{a_{HI}}}
\def\ahip{\ensuremath{\ahi^{\prime}}}
\def\pahip{\ensuremath{PA_{HI}^{\prime}}}

\def\apjl{{ApJL}}
\def\apj{{ApJ}}
\def\apjs{{ApJS}}
\def\aj{{AJ}}
\def\aap{{A\&A}}
\def\aaps{{A\&AS}}

\def\mnras{{MNRAS}}

\usepackage{epsfig}

\begin{document}

\title{Clouds Toward the Virgo Cluster Periphery: \\
Gas-rich Optically Inert Galaxies}
\author {
Brian R. Kent\altaffilmark{1,2}
}

\altaffiltext{1}{Jansky Fellow of the National Radio Astronomy Observatory.}

\altaffiltext{2}{National Radio Astronomy Observatory, 520 Edgemont Road, Charlottesville, VA 22903.
The National Radio Astronomy Observatory is a facility of the  
National Science Foundation operated under cooperative agreement by  
Associated Universities, Inc.
{\it e--mail:} bkent@nrao.edu}

\begin{abstract}
Aperture synthesis observations of two HI cloud complexes located in
the periphery of the Virgo galaxy cluster 
are presented. These low HI-mass clouds ($M_{HI}<$ 10$^{9}$) 
are seen projected on the M region of the western Virgo cluster, where the galaxy population is thought to lie behind the main A cluster surrounding M87.
The kinematic measurements of both  unresolved Arecibo and resolved VLA-C observations are in good agreement.
The HI detections cannot be identified with any optical, IR, or UV emission from available
archival imaging.
They are inert at these wavelengths.  The HI masses of the individual VLA detections range from 7.28 $\leq$~log($M_{HI}) \leq $ 7.85.  
The total dynamical mass estimates are several times their HI content, ranging from 7.00 $\leq$~log($M_{dyn}) \leq $ 9.07,
 with the assumption that the clouds are self-gravitating and in dynamical equilibrium.
We report the observed parameters derived from the VLA observations.  One of these HI clouds
appears to be the most isolated optically inert detection observed in the outer reaches
of Virgo.\\

Accepted for publication in The Astrophysical Journal.

\end{abstract}



\section{Introduction}
\label{intro}

Galaxies are characterized by their stellar content, morphology, environment, dust,
and neutral and molecular gas content.  The 21 cm line of neutral hydrogen (HI) continues
to play an important role in diagnosing the star formation potential
of a galaxy and any past dynamic interactions with a cluster environment
or neighboring galaxies.  Blind HI surveys can sample the gas-rich population
of the local Universe and search for low-mass, low-surface brightness systems.

The Arecibo Legacy Fast ALFA (ALFALFA; Giovanelli \etal 2005a) survey is providing
such a sample of gas-rich objects.  The project utilizes  the seven-element Arecibo L-band Feed Array (ALFA) receiver system to 
conduct a wide area extragalactic HI investigation with the 305 meter Arecibo reflector.
The survey improves over previous first generation surveys (HIPASS: Barnes \etal~2001, Meyer \etal~2004; Wong \etal~2006;  ADBS: Rosenberg \& Schneider 2002) in spectral and spatial resolution, providing 5~\kms~ resolution with a 3\arcmin~beam.  The 7000 sq. deg. of surveyed
sky covers a velocity range out to $cz_{\odot}\sim$18,000~\kms.  This includes interesting areas of the local supercluster such
as the Virgo cluster at $cz_{\odot}\sim$1100~\kms.  The first and second ALFALFA Virgo catalogs (Giovanelli \etal~2007, Kent \etal~2008)
comprise a complete mass limited dataset ($M_{HI} \geq 2 \times 10^{7}$~\msun~at Virgo).
Sources identified in Kent \etal (2007; 2009) comprise a sample of optically inert HI detections; these objects are important
in understanding the fate of galaxies in a cluster environment and its surrounding periphery.
Previous detections that cannot be correlated with optical counterparts have been associated with nearby
galaxy groups or nearby spiral galaxies harassed by the cluster environment (Ryder \etal~2001; Minchin~\etal~2005; Oosterloo \& van Gorkom 2005; Haynes~\etal~2007). 

Nearby galaxy cluster environments are of great interest to HI studies
as both the gravitational potential and intra-cluster medium (ICM) perturb
the gas structure and morphology of galaxies.  Ram pressure stripping and tidal
encounters at different cluster radii result in spiral galaxy deficiencies in HI
of varying degrees.  While three-dimensional paths of galaxies through the cluster are often
difficult to ascertain, the fingerprint of the cluster-galaxy interaction
is well studied through aperture synthesis observations;
high resolution studies
show that HI radii are smaller than their optical counterparts and match
predictions of theoretical studies (Giovanelli \& Haynes 1983).

Aperture synthesis observations have the ability to resolve higher sensitivity single dish survey detections.  Resolved HI observations reveal the truncation
of disks, tidal tails, and the disturbed morphology of late-type spiral galaxies (Cayatte \etal 1990,1994; Chung \etal 2006).  All
of these properties are indicators of galaxy-galaxy and galaxy-cluster interactions.
It has been shown that HI deficiency in late-type galaxies decreases with increasing cluster radius (Giovanelli \& Haynes 1985).
An important question of galaxy and cluster evolution that remains is:
What happens to the stripped gas in the cluster environment?  Will it be
destroyed by ablation and evaporate into the cluster halo?  Is it possible
that an isolated cloud can survive and re-initiate star formation?

Here we present HI aperture synthesis observations of two HI clouds
in the Virgo Cluster periphery, initially reported
in Kent \etal (2007).  These clouds, unresolved
by Arecibo, are resolved into separate clumps with VLA-C observations.
In section 2 we discuss the original Arecibo observations, data reduction, and detections.
In section 3 we describe the follow-up VLA observations, data reduction, and detections.
In section 4 we detail the environment and neighboring galaxies of the HI clouds.  
In section 5 we discuss possible cloud origins and compare to other gas-rich optically inert phenomena.
Section 6 summarizes the results of the study.


\section{Single Dish Detections}
\label{AO}

The HI clouds described here were detected by the ongoing ALFALFA survey.
The meridian transit observing strategy uses a sky drift mode with
a 100 MHz bandwidth and 4096 channels per polarization,
centered at 1385 MHz.  Each raw scan contains 14 spectra ( 7 beams $\times$ 2 linear polarizations per beam),
with a sampling rate of 1 Hz and spectral resolution of 24.4 kHz ($\delta V = 5.1$~\kms ~at the rest frequency of the 21cm HI line).
 The system temperatures of the ALFA receivers during the observations were in the range 
$26~\rm{K} < \rm{T}_{sys} < 30~\rm{K}$, yielding a root mean square (rms) noise of 
$\sigma_m=2.5\,$mJy/beam in channels with $\delta V=5.1$~\kms.
The flagging, calibration, and gridding of the data into cubes are described in detail by Giovanelli \etal (2005a),
Kent (2008) and Kent \etal (2009).  Table~\ref{AOobs} describes the parameters of the Arecibo observations.

\begin{figure}
\begin{center}
\includegraphics[width=6.5in]{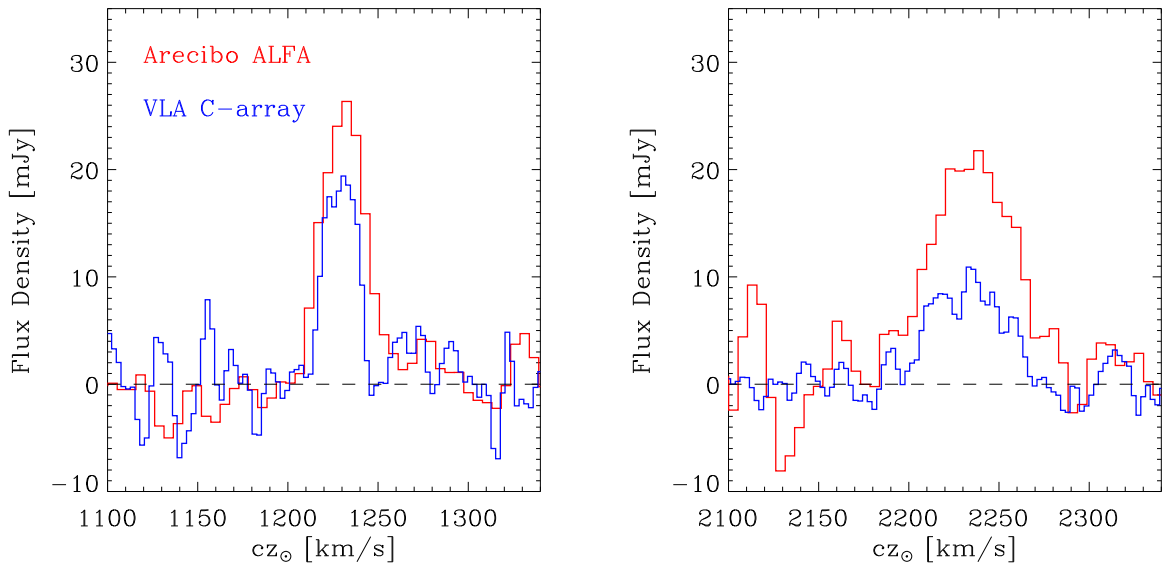}
\caption{Integrated spectral profiles of the Cloud 1 (left) and Cloud 2 (right) detections made with Arecibo and the ALFALFA survey (red),
and total integrated profiles from the multiple cloud detections in the same fields made with the VLA in C configuration (blue).
The channel resolution of the VLA spectrum is 12.2 kHz (2.6~\kms).  The global profiles for the VLA were obtained
by summing all emission for $\nhp \geq 1.5 \times 10^{20}$~\cm~in Figure~\ref{C1moments} and for $\nhp \geq 0.7 \times 10^{20}$~\cm~in Figure~\ref{C2moments}.\label{AOspectra}}
\end{center}
\end{figure}

The two HI clouds (henceforth Cloud 1 and 2) were detected in the ALFALFA data obtained in the
Spring 2005 campaign sampling the Virgo cluster and its surrounding periphery.
The detections were reported in Kent \etal (2007) as part of an initial
collection of gas-rich, optically inert extragalactic objects.  A complex
of HI clouds situated halfway between M87 and M49 and their VLA observations
were examined in Kent \etal (2009).  Here we continue this effort with two HI clouds 
located in the M cloud periphery west of the main A cluster.  
Upon detection in the ALFALFA survey, both objects
were re-observed and confirmed with the single pixel L-band Arecibo receiver.  
Integrated spectral profiles for each cloud are depicted in red in Figure~\ref{AOspectra}.
The Arecibo observations show both sources as narrow, single peaked spectral profiles.
Clouds 1 and 2 are unresolved by the Arecibo beam and 
are located 5.4$^{\circ}$ (1.5 Mpc in projection) and 4.2$^{\circ}$ (1.2 Mpc
in projection) from M87 respectively west toward the direction of the M cloud.
The Arecibo detections are unresolved point sources and we cannot deduce any information
about the morphology of the sources.

Table~\ref{AOparams} describes the observed parameters and locations of these two HI detections as measured from the ALFALFA data cubes, derived in the manner described by Giovanelli \etal (2007).
The spatial centroid of each cloud is in col.~(2). Its accuracy depends 
on the source strength, and is of order $\sim30\arcsec$~for the reported sources. The heliocentric velocity $cz_\odot$, 
width at 50\% of the peak $W_{50}$ and total flux $F_c$ of the integrated spectral profiles in 
Figure~\ref{AOspectra} are in cols.~(3)--(5). The signal-to-noise ratio $S/N$ of the 
detections is in col.~(6), and is given by
\be
        S/N=\Bigl({1000 \, F_c \over W_{50}}\Bigr){w^{1/2}_{smo}\over \sigma_{rms}} \;\;,
        \label{snr}
\ee
where $F_c$ is in Jy \kms, $W_{50}$ is in \kms, $w_{smo}$ is a smoothing width equal to the number of $10\,$ \kms~bins bridging half the signal, and $\sigma_{rms}$ is the rms noise (in mJy) across the integrated spectrum at 10~\kms\ resolution. The HI mass \mhi~ for each cloud is in col.~(7), and is computed assuming that the clouds are optically thin:
\be
M_{HI}/M_\odot = 2.356 \times 10^5 \, D^{2} \, F_c \;\;,
\label{HImass}
\ee        
where $D$ is in Mpc and \fc\ is in Jy~\kms (Roberts~1975). The uncertainties on \mhi\ in Table~\ref{AOparams} and elsewhere do not include errors in the distance adopted, which is poorly constrained due to the large peculiar velocities of objects near or within the cluster.
As described in Giovanelli \etal (2005b) and Springob \etal (2006a), this results in ambiguities for galaxies with $cz_{\odot}<3000$~\kms.
The peculiar flow model used for the ALFALFA distances published in the catalogs corrects only for large-scale perturbations in the velocity field.
The model is not able to deal effectively with regions in the immediate vicinity of Virgo.  We adopt the same distance values for consistency with Kent \etal (2007):
16.7 Mpc for Cloud 1 and 34.8 Mpc for Cloud 2.  The model used to obtain the distances is based on the work of Tonry \etal (2000) and Masters \etal (2004),
using a parametric model and spherical truncated power-law attractor to examine the peculiar motions that arise from a cluster like Virgo.

\section{Aperture Synthesis Follow-up Observations}
\label{VLA}

Both HI cloud centroid positions were observed with the Very Large Array\footnote{The VLA is a 
facility of National Radio Astronomy Observatory, which is operated by Associated Universities, Inc., 
under a cooperative agreement with the National Science Foundation.} in November 2006.
Approximately nine hours of on-source integration were obtained in C configuration for each source.
Online Hanning smoothing yielded a channel spacing of 12.2 kHz over a bandpass of 1.5 MHz.

The data from the runs were reduced using the Astronomical Image Processing System 
({AIPS}; Greisen 2003) as described in Kent \etal (2009).
Standard flux, phase and bandpass calibration and continuum subtraction routines were applied after flagging.
The calibrated data were imaged using various weighting schemes;
we analyze the highest sensitivity, naturally-weighted cubes with a 
synthesized beam width of $\sim$25\arcsec~($\sim$2.0 kpc at the Virgo distance).  
The data cubes created for the Cloud 1 field are not limited by dynamic range, and do not gain image fidelity from self-calibration.
For the field with Cloud 2, self calibration
was run using a strong continuum source (NVSS catalog position $\alpha=$12$^h$13$^m$32.1$^s$, $\delta=$+13$^{\circ}$07\arcmin20.4\arcsec; Condon \etal~1998) of flux density 1.3 Jy,
greatly improving the fidelity and phase calibration of the images.
As part of the reduction process, each image was smoothed to the resolution of the ALFALFA data cubes
to identify emission in channels, as well as their extent in the frequency domain.
A summary of the aperture synthesis observing and map parameters is given in Table~\ref{VLAobs}.
For clarity, all variables denoting parameters derived from the VLA observations are primed.

\begin{figure}
\epsscale{1.0}
\plotone{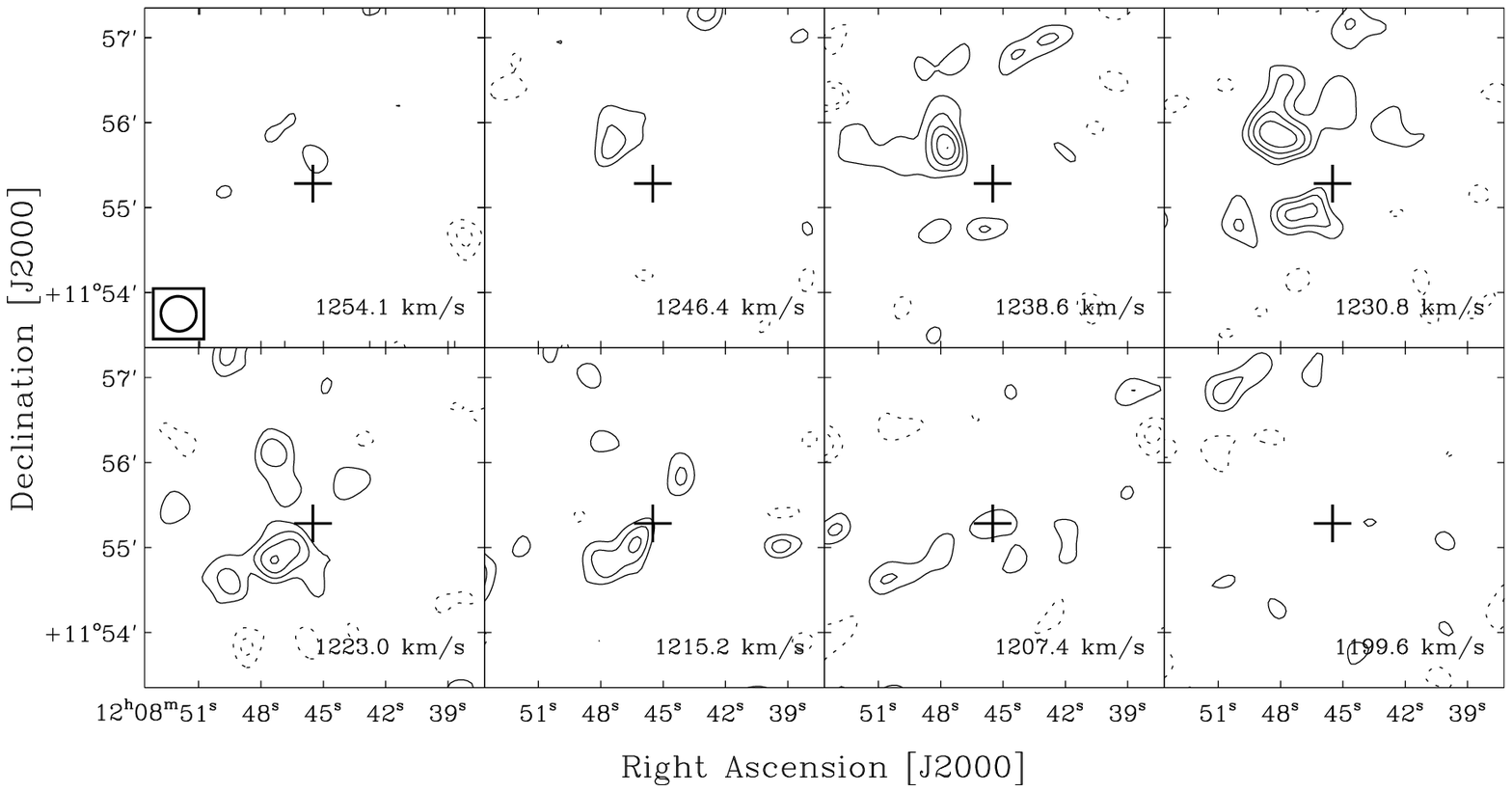}
\caption{Naturally-weighted channel maps for the Cloud 1 field from the VLA-C observations. The plotted channels are independent ($\chanp = 7.8~\kms$). Contours are at (-3, -2, 2 ($2\sigma_{m}^{\prime}$), 3, 4, 5, 6) mJy/beam; 
negative contours are represented with dashed lines. The cross indicates the centroid position of the original Cloud 1 Arecibo detection
(Table~\ref{VLAparams}).  The heliocentric radial velocity is in the lower right corner of each panel, and the synthesized beam is in the lower left corner of the first panel. 
\label{C1chans}}
\end{figure}

\begin{figure}
\epsscale{1.0}
\plotone{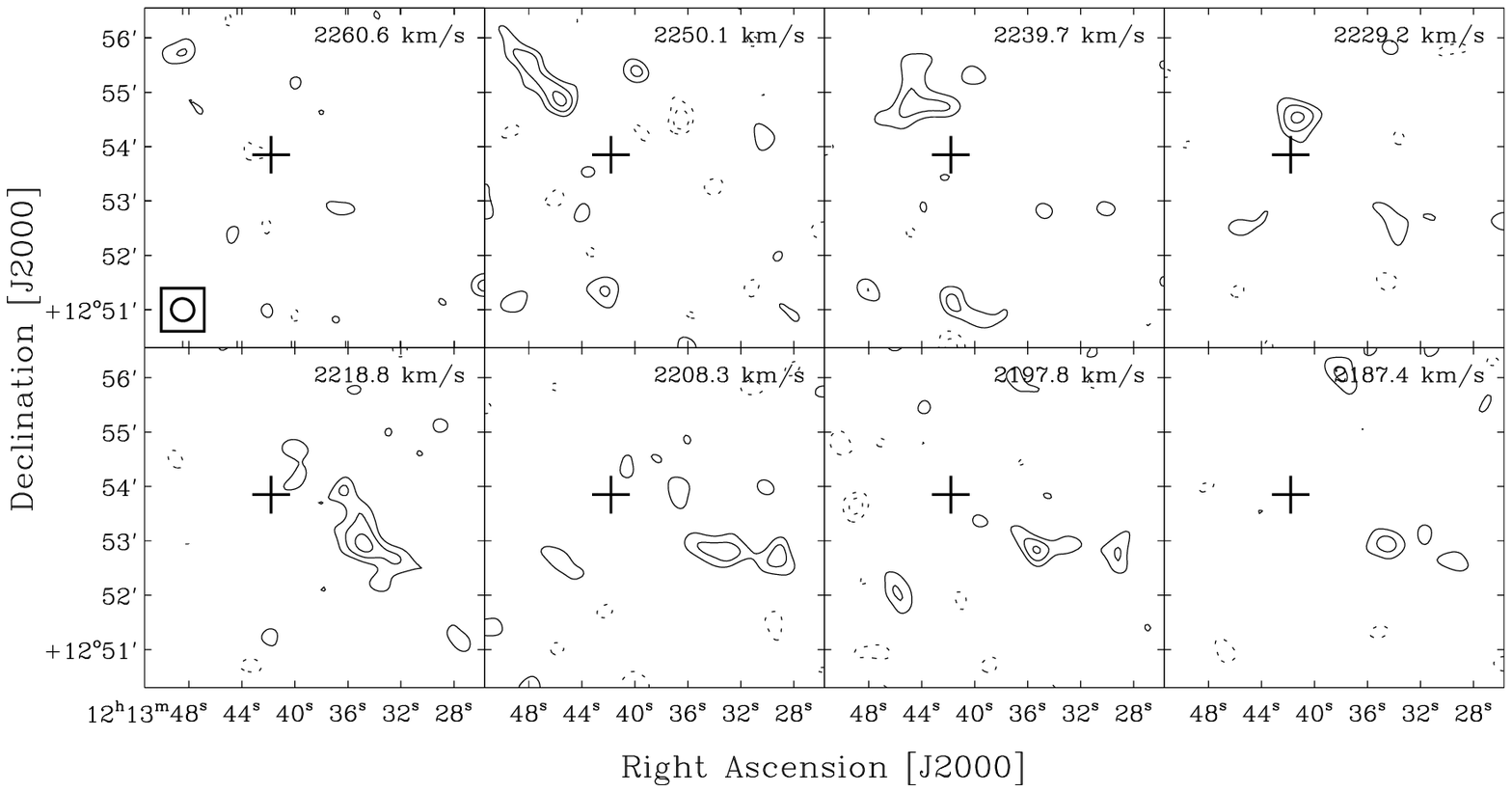}
\caption{Naturally-weighted channel maps for the Cloud 2 field from the VLA-C observations. The plotted channels are independent ($\chanp = 10.4~\kms$). Contours are at (-3, -2, 2 ($2\sigma_{m}^{\prime}$), 3, 4, 5, 6) mJy/beam; 
negative contours are represented with dashed lines. The cross indicates the centroid position of the original Cloud 2 Arecibo detection
(Table~\ref{VLAparams}).  The heliocentric radial velocity is in the upper right corner of each panel, and the synthesized beam is in the lower left corner of the first panel. 
\label{C2chans}}
\end{figure}

Channel maps of each field are shown in Figure~\ref{C1chans} and~\ref{C2chans},
with solid and dashed contours in the primary beam--corrected maps depicting, respectively,
positive and negative multiples of the median rms map noise.
All maps are corrected for the attenuation of the 
primary beam, and averaged over 3-4 spectral channels to yield a channel map resolution of 
$\delta V^{\prime} = 7.8$~or~$10.4$~\kms.
The emission from detections in both fields is contiguous over multiple channels in different
weighting and imaging deconvolution schemes.

Total intensity (zero moment) and intensity-weighted velocity (first moment) maps of each field are shown in 
Figures~\ref{C1moments} and \ref{C2moments}.  The zero moment contours are overlayed on SDSS $g$-band images.
For each frequency channel we blank regions with less than
3\smp\ (defined in Table~\ref{VLAobs}) in the image before combining into the zero moment map.
The first moment maps are computed only at locations with column densities of 
\nhp$\geq 1.5 \times 10^{20}$ cm$^{-2}$ (Cloud 1) and \nhp$\geq 0.7 \times 10^{20}$ cm$^{-2}$ (Cloud 2).
Global integrated spectral profiles from $all$ detections
in each field, representing the total emission for comparison with the original
ALFALFA spectrum are shown in Figure~\ref{AOspectra} in blue.
The rms error on the computed total emission over the full width half max (FWHM) range in each individual channel for both spectra range from
0.67 to 1.03 mJy and also reflects a $5\%$ uncertainty in calibration.

\subsection{HI Aperture Synthesis Detections}

\begin{figure}
\epsscale{1.0}
\plotone{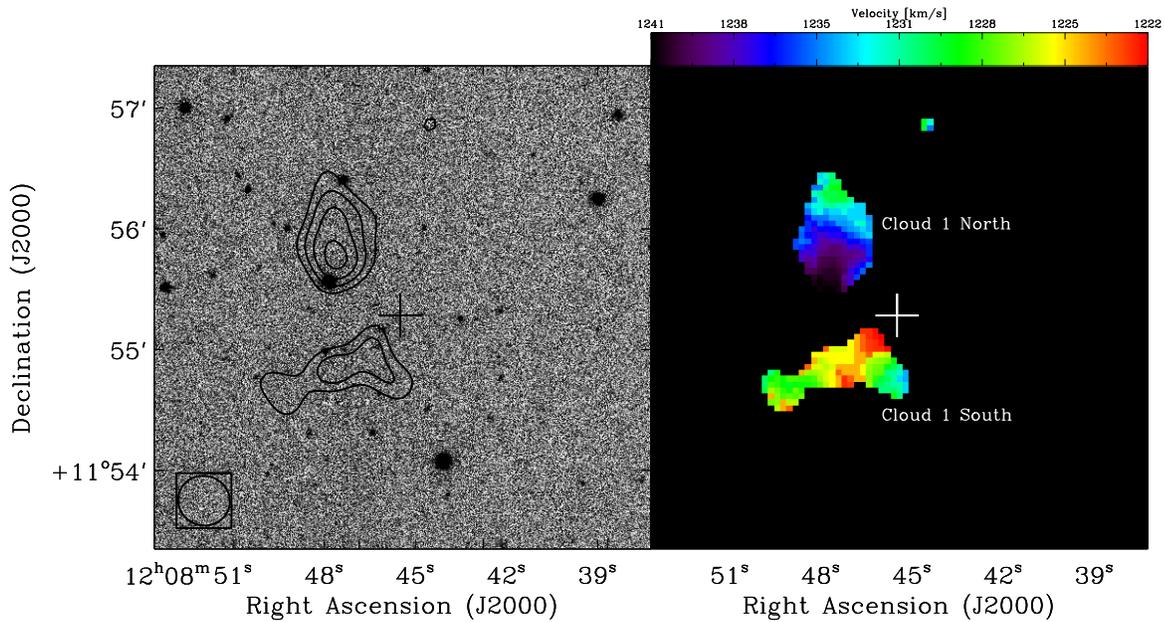}
\caption{\hi\ distribution and kinematics of the Cloud 1 Field, showing Cloud 1 North and 1 South as detected in the VLA-C data cube. 
The left panel shows total intensity map of the clouds (contours) superimposed on an SDSS $g$-band image (grayscale). Contours are at $\nhp=10^{20} \times\,$(1.5, 2, 2.5, 3)~\cm, and the grayscale is plotted logarithmically. The cross indicates the centroid position
of the original Cloud 1 Arecibo detection (Table~\ref{VLAparams}).  The synthesized beam is in the lower left corner of the panel. The right panel shows intensity-weighted velocity map of the clouds in regions where $\nhp \geq 1.5 \times 10^{20}~\cm$. The velocity spans 1222--1241~\kms\ on a linear scale, as indicated by the colorbar at the top of the plot.
\label{C1moments}}
\end{figure}

\textbf{Field for Cloud 1:}  We observe two detections which we label Cloud 1N (North) and 1S (South),
the two brightest in the field.  Individual integrated spectral profiles of each detection
are shown in Figure~\ref{C1moments}.  The northern cloud 1N appears to be more aligned north--
south whereas the lower signal-to-noise southern cloud 1S is aligned east--west.  
The northern cloud 1N exhibits a clear north--south velocity gradient.
The velocity gradient of the southern cloud 1S is slightly more disordered
but trends along the east/west axis of the detection.  
There is no discernible structure on smaller scales or clumpiness as seen in the Virgo HI objects
examined in Kent \etal (2009).  No counterparts
are observed in the field in any available optical, IR, or UV imaging databases.
The combined kinematic properties of Clouds 1N and 1S agree well with the Arecibo data (Tables~\ref{AOparams}~and~\ref{VLAparams}).

We recover $87 \pm$ 5\% of the flux (Tables~\ref{AOparams} and~\ref{VLAparams}) from the Cloud 1 field in the VLA observations.
The VLA global profile of all the integrated emission (which consists of Cloud 1N and 1S) for $\nhp \geq 1.5 \times 10^{20}~\cm$
in Figure~\ref{C1moments}~is shown in comparison with the Arecibo spectra in Figure~\ref{AOspectra}.
The mean velocity of Clouds 1N and 1S is 1229.5 \kms~and agrees with the centroid velocity of the 
Arecibo data.  Some lower surface brightness HI might escape detection below the 3$\sigma^{\prime}_m$
level if it is distributed uniformly over the 30\arcsec~region between the two detections.  However,
one would expect it to coincide kinematically with the mean velocity of the clouds.

Comparing the total HI mass of Clouds 1N and 1S from the VLA with that 
of Arecibo reveals little deficiency in the HI within the errors of the measurement
(Tables~\ref{AOparams} and~\ref{VLAparams}).  No detailed morphology can 
be ascertained from the VLA observations at the achieved sensitivity and spatial resolution, i.e.
within each of the Clouds 1N and 1S.  This suggests that no flux exists on scales greater than $\sim$1\arcmin~
and that any detectable structure of the detected blobs will be smaller than that spatial scale. 

\textbf{Field for Cloud 2:}  The emission detected in this field is, not unexpectedly, of
lower signal-to-noise. The VLA global profile of all the integrated emission for $\nhp \geq 0.7 \times 10^{20}~\cm$~
in Figure~\ref{C2moments}~is shown in comparison with the Arecibo spectra in Figure~\ref{AOspectra}.
We recover $41 \pm$ 5\% of the flux (Tables~\ref{AOparams} and~\ref{VLAparams}) compared
to the Arecibo data.  It is likely that lower column density emission is resolved out
and we are only detecting the higher column density peaks of the source.  We identify
three of these peaks (Cloud 2 North, West, and South) fit for the same parameters as the Cloud 1 field.
Clouds 2 North and West are elongated in an east-west direction, where the Southern component
is a marginal detection in the map with low signal to noise, albeit still a 5$\sigma^{\prime}_m$ detection
in the map.  The velocity maps of all three Cloud 2 field detections show no ordered motion.
The clutter in the field contains a number of bumps one or two sigma above the rms noise.

\subsection{Field Properties}

\begin{figure}
\epsscale{1.0}
\plotone{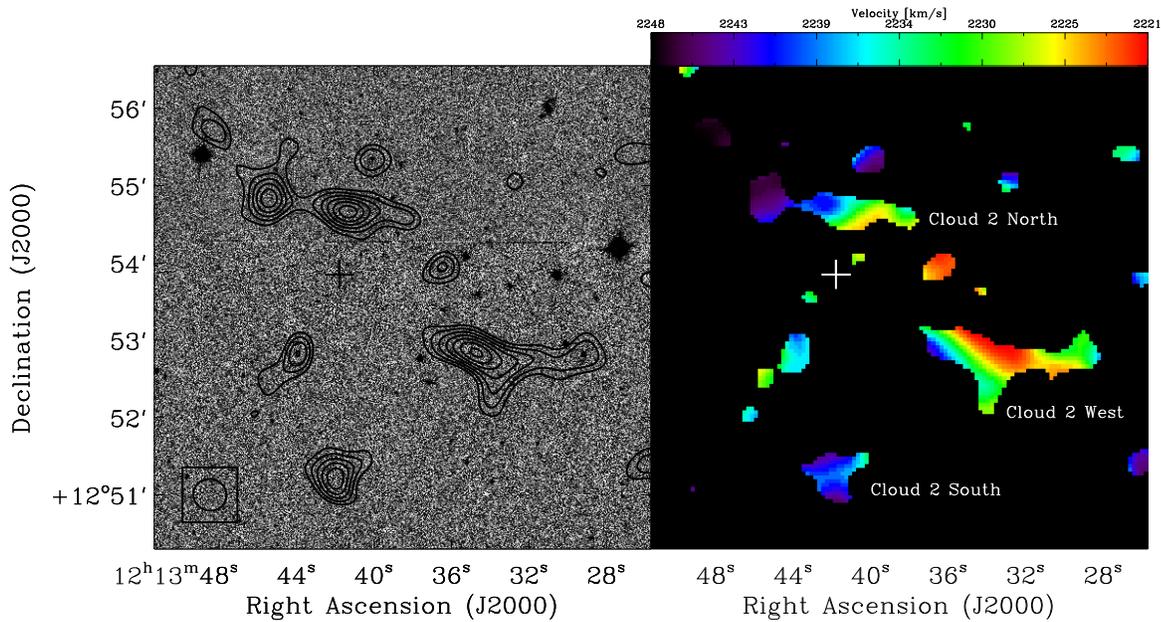}
\caption{\hi\ distribution and kinematics of the Cloud 2 Field. 
The left panel shows total intensity map of the clouds (contours) superimposed on an SDSS $g$-band image (grayscale). Contours are at $\nhp=10^{20} \times\,$(0.7,0.8,0.9,1.0,1.1,1.2,1.3,1.4,1.5)~\cm, and the grayscale is plotted logarithmically. The cross indicates the centroid position
of the original Cloud 2 Arecibo detection (Table~\ref{VLAparams}).  The synthesized beam is in the lower left corner of the panel. The right panel shows intensity-weighted velocity map of the clouds in regions where $\nhp \geq 1.5 \times 10^{20}~\cm$. The velocity spans 2222--2248~\kms\ on a linear scale, as indicated by the colorbar at the top of the plot.
\label{C2moments}}
\end{figure}

\begin{figure}
\epsscale{1.0}
\plotone{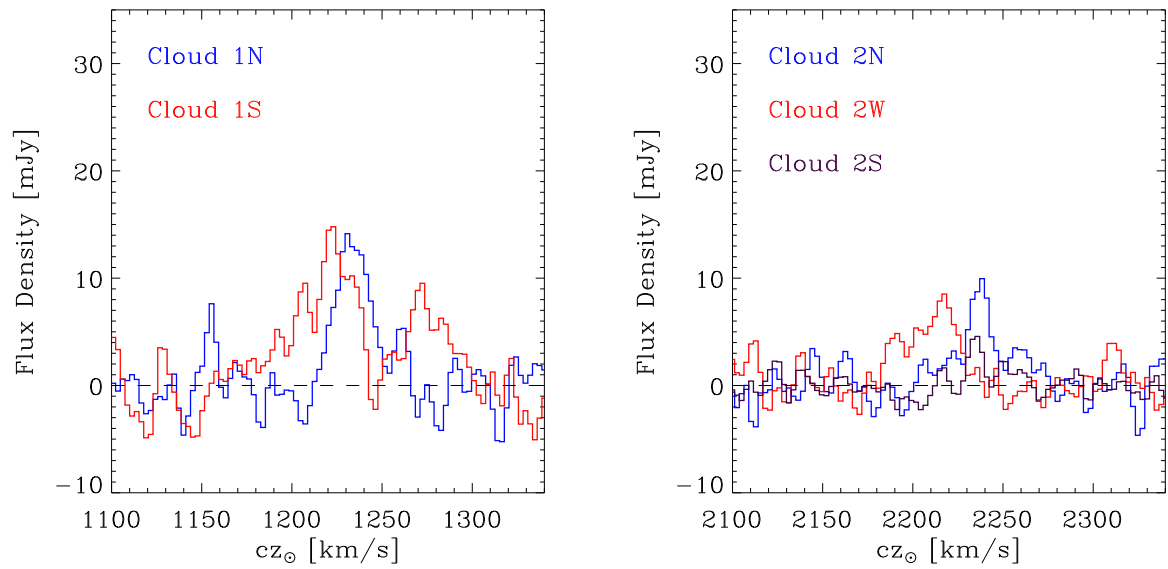}
\caption{Integrated spectral profiles of VLA-C detections in the Cloud 1 field (left) and Cloud 2 field (right).
\label{VLAspectra}}
\end{figure}

The properties for the detections in both fields are summarized in Table~\ref{VLAparams} in the same
manner as Kent \etal (2009).
The global emission parameters, where measurable, are also listed.
Each detected feature in the VLA data cubes was fit with a centroid ellipse in the same fashion as
the Arecibo data.
The centroid positions of these fits for each individual cloud detection in the total intensity maps
of Field 1 and 2 (Figures~\ref{C1moments} and~\ref{C2moments}) are given in col.~(2).
The centroid $cz_{\odot}^{\prime}$ of both the individual
integrated profiles of Figure~\ref{VLAspectra}~is in col.~(3), and $W_{50}^{\prime}$ of 
the profiles is in col.~(4). The values of $W_{50}^{\prime}$ are corrected for instrumental 
effects by assuming that the unbroadened profile is Gaussian. The integrated flux 
density \fcp~and HI mass $M_{HI}^{\prime}$ are in cols.~(5)~and~(8), respectively.  
The major axis $a_{HI}^{\prime}$ measured from the ellipsoidal fits of each individual cloud is in col.~(6). We adopt the 
outermost locations as the edges of the clouds where \nhp $= 1.5 \times 10^{20}$ 
cm$^{-2}$ (Figure~\ref{C1moments}) and~\nhp $= 0.7 \times 10^{20}$ 
cm$^{-2}$ (Figure~\ref{C2moments}).
The position angle \pahip\ at which \ahip\ is measured is in col.~(7). An estimate of the dynamical mass 
$M_{dyn}'$ of each cloud is in col.~(9), and is computed via:
 \be
 M_{dyn}' = (3.39 \times 10^4)\,a_{HI}^\prime D \left( \frac{W_{50}^{\prime}}{2} \right)^2 \;\;,
 \label{Dynmass}
 \ee 
where $a_{HI}^{\prime}$ is the object diameter in arcminutes, $W_{50}^{\prime}$ is in \kms\ and the 
distance $D$~is in Mpc.  We note that $M_{dyn}'$ has physical meaning $only$ 
if the clouds are self-gravitating and in dynamical equilibrium, which may or may not
be a valid assumption.

 
\section {The Environment of the Cloud Complexes}\label{optenv}

\begin{figure}
\begin{center}
\includegraphics[width=6.5in]{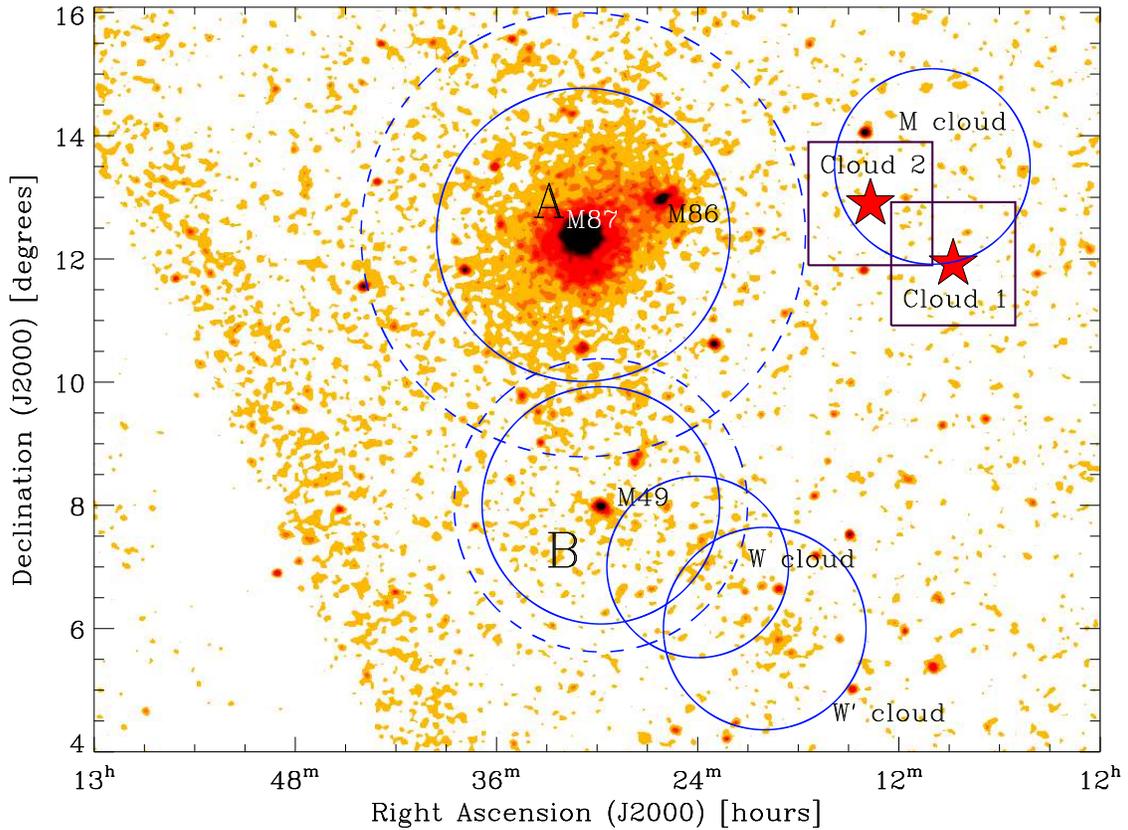}
\caption{The location of the two Arecibo HI cloud detections in the context of the greater cluster environment.
	  The centroid positions of Cloud 1 and Cloud 2 are indicated by star symbols.
          The peaks from the hard X-ray background image provided by ROSAT are labeled
          indicating Virgo cluster galaxies M49 and M87 (Snowden \etal~1995).  
	  The symbols are not indicative of source sizes and 
          are shown only for positional indication.  The dashed lines indicate the projected 
	virial radius of the dark matter halo $r_{200}/2$ determined by McLaughlin (1999),
	for the A and B areas centered around cluster members M87 and M49 respectively. 
	The two 2$^{\circ}$ boxes surrounding the detections show the areas of sky examined in Figure~\ref{skyarea}.
\label{virgoenviron}}
\end{center}
\end{figure}

Figure~\ref{virgoenviron} shows
the location of these two HI clouds with respect to the Virgo Cluster, with the
boundaries and areas of Binggeli \etal (1993) and plotted against
the hot X-ray cluster background detected by ROSAT (Snowden \etal 1995).
Both cloud regions are also outside the projected virial radius of the dark
matter halo around M87 determined by McLaughlin (1999).
The detections lie in the vicinity of the M cloud (Ftaclas \etal 1984)
west of M87, where member galaxies are considered
to lie behind the main A cluster.  These M cloud
galaxies have a larger mean velocity ($cz_{\odot}\sim2000$~\kms) than
the main cluster ($cz_{\odot}\sim1150$~\kms).  However, peculiar velocities
due to the large central mass of the cluster yield a velocity
dispersion of the projected M cloud region galaxies that ranges from -100 $ < cz_{\odot} < $ 2400~\kms.
The assignment of these HI features to the various areas of Virgo remains ambiguous.


Imaging studies by Roberts \etal (2007) analyzed an area extending
from the eastern part of the M cloud northward.  They arrived at a density of  20 to 60 low surface brightness (LSB) dwarf galaxies per
square degree in the area 1.5$^{\circ}$ west of M87.  This may have some relation
to possible parent galaxies of the Cloud 2 field; Figure~\ref{virgoenviron}~shows that the Cloud 1 field 
is more removed from the main cluster and this density measurement.

\begin{figure}
\begin{center}
\includegraphics[width=6.5in]{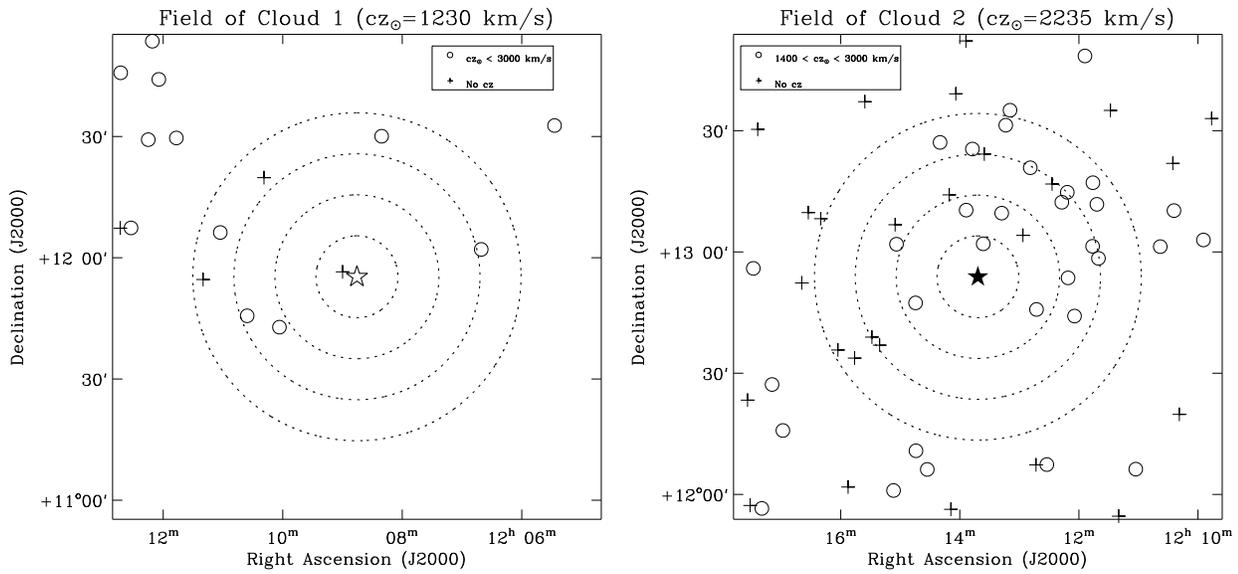}
\caption{Galaxies in the vicinity of the two cloud complexes.
         The plots show 2$^{\circ} \times 2^{\circ}$ areas of sky surrounding the Arecibo detection Cloud 1 (left; white star),
	 and Cloud 2 (right; dark star).
	 The plot shows all objects with published redshifts of $cz_{\odot} <$ 3000~\kms~for the Cloud 1 area
         and 1400 $<~cz_{\odot} <$ 3000~\kms~for the Cloud 2 area.
         The crosses indicate galaxies without published redshifts.
	 The dashed circles are of radii 50, 100, 150 and 200 kpc projected on the sky at the Virgo Cluster distance
         of 16.7 Mpc.\label{skyarea}}
\end{center}
\end{figure}

No obvious optical features that resemble an extragalactic counterpart
can be correlated with any of the cloud components
in online imaging databases.
The catalogs provided by the NASA/IPAC Extragalactic Database (NED), Sloan Digital Sky Survey (York \etal 2000),
Virgo Cluster Catalog (Binggeli \etal~1985), GOLDMine (Gavazzi \etal~2003) and published
ALFALFA survey (Giovanelli \etal~2008; Kent \etal~2009) were
examined for possible nearby associations to the Cloud 1 and 2 fields.
Figure~\ref{skyarea} shows 2$^{\circ} \times 2^{\circ}$ areas of sky surrounding
each of the Arecibo detections.  Each plot contains open circles from galaxies with
published HI or optical redshifts within a given range of the measured Arecibo/HI velocity for
the detection; $cz_{\odot} \leq$ 3000~\kms~for the Cloud 1 area
and 1400 $<~cz_{\odot}~<$ 3000~\kms~for the Cloud 2 area.  Galaxies in those
areas with no published redshifts at any wavelength are depicted as small crosses.
We next examine a number of nearby galaxies of comparable redshift within the projected vicinity of Cloud 1
and Cloud 2.

\subsection{The Projected Environment of Cloud 1}

The galaxy environment below a redshift of $cz_{\odot} \leq 3000$~\kms~is rather
	sparse in the vicinity of Cloud 1.  Galaxies that meet this criteria
	are listed in Table~\ref{Cloud1environ}.  Of particular note
	is the faint galaxy SDSS J120859.92+115631.2.  This faint detection lies 3.8\arcmin~NE of the Arecibo
	centroid.  It is the closest optical detection near Cloud 1; 
	it remains ambiguous as to whether this object is a Virgo Cluster member 
	or a more distance background galaxy.  The SBb(r)I-II galaxy VCC 58
	has a disturbed morphology and optical redshift of 
 	$cz_{\odot}=2188$~\kms (Rubin \etal 1999) and HI redshift
 	of $cz_{\odot}=2209$~\kms (Giovanelli \etal 2007).  
	This is the only spiral galaxy within $\sim 1^{\circ}$ though its velocity differs
	by $\sim$1000~\kms. It is described as having a disturbed rotation curve (Rubin \etal 1999).

\subsection{The Projected Environment of Cloud 2}

Cloud 2 lies in a much higher galaxy density environment and includes, within 2$^{\circ}$, nine galaxies with a cataloged late type morphology for 1400 $<~cz_{\odot}~<$ 3000~\kms (Table~\ref{Cloud2environ}).  Several galaxies 
have notable relevant properties.  The aforementioned VCC 58 lies 49\arcmin~southwest of Cloud 2
and is of comparable redshift.  The closest published HI detection is VCC 85,
 detected by Gavazzi \etal (2006) with $cz_{\odot}=1932$~\kms and
                                           lies $\sim$8\arcmin~from the Cloud 2 centroid.
Also of note is VCC 97, an SAB galaxy that lies 17\arcmin~ North of Cloud 2.	
					   Chamaraux \etal~(1980) showed this galaxy to be HI deficient 
					   (DEF$_{HI}$=0.21; Helou, Hoffmann, \& Salpeter 1984); 
					   it has an HI redshift of $cz_{\odot}=$2470~\kms.  
					   Doyon \& Joseph (1989) also noted a dust deficiency in VCC 97.

\section{Discussion}
\label{discussion}

Roberts (1988; see references therein) outlined various categories of intergalactic HI clouds:
tidal tails, extended HI envelopes, and clouds near groups or within clusters.  Starless
intergalactic clouds in the field, isolated from other galaxies, have yet to be detected.
Recent studies have focused on the search and identification
of starless gas-rich halos.  As one of the many important science goals, 
identifying such objects in blind surveys like HIPASS and ALFALFA
give useful information on the formation and evolution of galaxies
in a variety of environments.  Recently detected optically inert clouds and their respective
followup studies can be associated with galaxies in nearby clusters 
(Sancisi \etal~1987; Davies \etal~2004; Minchin \etal~2005; Haynes \etal~2007; Kent \etal~2007),
in groups or disturbed galaxies (Schneider \etal~1983; Henning \etal~1993; Ryder \etal~2001),
in tidal or harassed tails (Oosterloo \& van Gorkom 2005; Giovanelli \& Haynes 1989; Salzer \etal 1991),
or as an HVC or Milky Way/Local Group companion (Kilborn \etal~2000; Giovanelli \etal~2010).
However, surveys have not revealed a large population of previously undetected dark matter halos
predicted by large scale simulations (Moore \etal~1999).  In the nearby Virgo Cluster, 
the HI detected in clouds or tidal streams does not make up a significant portion of the HI deficiency
in nearby parent spirals; the population of HI clouds does not, by itself, offer a complete
solution to the missing satellite or mass problem (Kent \etal~2009; Klypin \etal~1999).

None of the detections discussed here appear to be tidal tails that clearly extend to an obvious parent
galaxy.  The largest tail extending from a Virgo galaxy is near NGC 4532 at a length of 500 kpc (Koopmann \etal~2008).
While late type spirals and dwarfs are within a projected 500 kpc range of both Cloud 1 and Cloud 2, neither
has a tail or streamlike morphology leading to another nearby galaxy.
The clouds do not belong to a compact group of galaxies, nor are they are part of the main
A or B clusters surrounding M87 or M49.

Although the region surrounding the Cloud 1 detection
lies at a projected distance of 1.5 Mpc from M87, it has been shown that the spiral 
galaxy population in the M cloud area is HI deficient.
The Cloud 1 detection lies on the boundary of higher HI deficiency (Solanes \etal~2001),
whereas the Cloud 2 field lies within it.
The intracluster X-ray density in the vicinity of the Virgo M cloud is estimated to be
$n_{icm} \sim 3 \times 10^{-6}$ cm$^{-3}$ (computed from Vollmer \etal~2001).
The Virgo Cluster ICM temperature maps computed by Shibata \etal~(2001) do not
cover the region of sky containing the HI clouds.
If we entertain the assumption that these clouds came
from a spiral disk, then the presence of this gas deficiency means that a ram pressure stripping
hypothesis cannot be completely discarded.

We can place upper limits on the optical surface brightness based on models of Bell \etal (2003).
As in Kent \etal (2009) we assume a $g$--band imaging 
surface brightness limit similar to other SDSS LSB galaxy studies ($\mu_{g}\sim$26 mag arcsec$^{-2}$; Kniazev \etal 2004).
A feature of source size $\sim 10^{\prime\prime}$ would have a $g$--band luminosity of $L_{g}\sim 10^{6}~L_{\odot}$
and model stellar M/L ratio of $M^*/L^*\sim$1.6.  The theoretical upper limits
for the stellar to HI mass ratio would range from $\sim$0.02 to 0.11 for the clouds extracted from the VLA data cubes.
Upper limits for the HI mass to stellar luminosity would range from $\sim$15 to 70.  
It remains an open issue as to whether or not
any optical emission can be positively correlated with these HI detections.

As indicated in Section~\ref{optenv}, nine late-type galaxies lie in the vicinity and near redshift range of Cloud 2; only
one lies $\sim$1$^{\circ}$ Northeast of Cloud 1.   This makes it difficult to identify a parent galaxy.
However, we can hypothetically consider the movement of these clouds through the cluster environment.
As both clouds are at higher velocities than the systemic heliocentric cluster
velocity ($cz_{\odot,Virgo}\sim$1150;~\kms~Huchra 1988), their line-of-sight velocity
with respect to the cluster reference frame is directed away from us.
If the clouds were torn from a spiral disk that is in a similar reference
frame, with the clouds decelerating, then the parent galaxy would be at a higher systemic
velocity than the cluster.  
The only nearby spiral galaxy of comparable velocity is the aforementioned VCC 58 (IC 769), 
located one degree
Northeast of the Cloud 1 detection at a redshift $cz_{\odot}$=2209~\kms.
VCC 58 also lies one degree
Southwest of Cloud 2, and stands as a remote, yet possible candidate parent of either cloud.

\section{Summary}
\label{summary}

We have presented new follow-up observations obtained with  
the Very Large Array that resolve original Arecibo HI detections of extragalactic
HI clouds in the Virgo Cluster periphery.
The results of these observations are summarized as follows:

\begin{enumerate}
\item Two HI clouds detected and unresolved with Arecibo using ALFALFA survey data. 
	The HI detections have heliocentric radial velocities of $cz_{\odot}=$ 1230
	and 2235~\kms.  The velocity widths are narrow at 29 and 53~\kms.  The HI masses
	of Cloud 1 and 2 are, respectively, 4.3$\times 10^{7}$ and 3.5$\times 10^{8}$\msun.

\item Detections have been made with the VLA in both the Cloud 1 and Cloud 2 fields
	at the same velocities as the Arecibo detections.  The data show
	two and three separated regions of HI emission for the Cloud 1 and Cloud 2 fields respectively.
	The individual HI masses range from log($M_{HI}^{\prime}$)=7.1--7.8~\msun.  We recover 87\% of
	the flux for the Cloud 1 field and 41\% of the flux for the Cloud 2 field.
	No optical, IR, or UV counterpart can be identified with these HI features using 
	available online imaging databases.

\item The galaxy environment is relatively sparse around Cloud 1 - one faint object with no redshift information, 
        SDSS J120859.92+115631.2, lies 3.8\arcmin~northeast of the Arecibo centroid.
        The nearest late-type galaxy of comparable Virgo redshift is VCC 58, located one degree to the northeast.

\item The Cloud 2 detection lies in a dense galaxy environment showing higher HI deficiency with nine late type 	
	spiral systems of comparable Virgo redshift within a one degree radius.  The closest HI detection is VCC 85
	at 8\arcmin.

\item The HI deficient spirals in the M cloud region show that 
        dynamic processes are prevalent even at large distances from the Virgo Cluster center.
        While there are no larger spirals immediately in the vicinity or at comparable velocity 
        of the HI Clouds, we cannot dismiss a cloud origin hypothesis of ram pressure stripping.
	Much like previous detections reported in Kent \etal (2007; 2009), 
        it is unlikely that the HI clouds described here are primordial gas structures in dark matter halos.
	These two clouds are located in the outer parts of the cluster and are in a lower density environment
	than other HI clouds and tidal tails further toward M87 or M49.
	Cloud 1 remains unique in its isolation.  To date, there are no other gas structures
	that are both definitively extragalactic and unambiguously not associated with another galaxy
	outside the Local Group.

\end{enumerate}

We wish to thank David E. Hogg and Morton S. Roberts of the NRAO for their
assistance and encouragement of this work.  We also wish to thank William Cotton
for advice on imaging techniques, and the anonymous referee for their careful review.\\

{\footnotesize

BRK acknowledges support from a Jansky Fellowship during the completion of this work.
This research has made use of the NASA/IPAC Extragalactic Database (NED) which is 
operated by the Jet Propulsion Laboratory, California Institute of Technology, 
under contract with the National Aeronautics and Space Administration.   $Skyview$ was developed 
and maintained under NASA ADP Grant NAS5-32068
under the auspices of the High Energy Astrophysics Science Archive Research Center at the 
Goddard Space Flight Center Laboratory of NASA.  

This research has made use of Sloan Digital Sky Survey (SDSS) data. Funding for the SDSS 
has been provided by the Alfred P. Sloan Foundation, the Participating Institutions, the National Aeronautics 
and Space Administration, the National Science Foundation, the U.S. Department of Energy, the Japanese Monbukagakusho, 
and the Max Planck Society. The SDSS Web site is http://www.sdss.org/.  The SDSS is managed by the Astrophysical Research 
Consortium (ARC) for the Participating Institutions. The Participating Institutions are The University of Chicago, Fermilab, 
the Institute for Advanced Study, the Japan Participation Group, The Johns Hopkins University, the Korean Scientist Group, Los
Alamos National Laboratory, the Max-Planck-Institute for Astronomy (MPIA), the Max-Planck-Institute for Astrophysics (MPA), 
New Mexico State University, University of Pittsburgh, University of Portsmouth, Princeton University, the United States 
Naval Observatory, and the University of Washington.

}

\newpage

\clearpage

\onecolumn

\twocolumn

\newpage

\begin{deluxetable}{lc}
\tablecaption{ALFALFA Observing and Data Cube Parameters \label{AOobs}}
\tablewidth{0pt}
\tablehead{
	 \colhead{Parameter} & 
	 \colhead{Value}
}
\startdata
Spectral range	                          &      25 MHz (-2000 -- 3200 \kms) \\
Effective integration time                         &      48 seconds (beam solid angle)$^{-1}$\\
Spectral resolution \chan\                 &      24.4 kHz (5.1 \kms)\\
Half-power beam size                               &      $3\arcmin.3\ \times 3\arcmin.8$ \\
RMS noise $\sm$ for $\chan = 5.1\,$\kms           &      2.5 mJy/beam\\

\enddata
\end{deluxetable}

\begin{deluxetable}{ccccccc}
\tablecaption{Arecibo Single-Dish Cloud Properties\label{AOparams}}
\tablewidth{0pt}
\tabletypesize{\footnotesize}
\tablehead{
            \colhead{Cloud} & 
	    \colhead{$\alpha,\delta$} & 
	    \colhead{\czsun} & 
	    \colhead{\wf} & 
	    \colhead{\fc} &
	    \colhead{\sn} &
	    \colhead{$\log(\mhi/\msun)$}
	    \\ 	    
	    \colhead{ } & 
	    \colhead{(J2000)} & 
	    \colhead{(\kms)} & 
	    \colhead{(\kms)} & 
	    \colhead{(Jy \kms)}  & 
	    \colhead{ } & 
	    \colhead{ } 
	    \\	    
	    \colhead{(1)} & 
	    \colhead{(2)} & 
	    \colhead{(3)} & 
	    \colhead{(4)} & 
	    \colhead{(5)} & 
	    \colhead{(6)} &
	    \colhead{(7)} 
}
\startdata
Cloud 1  & $12\,08\,45.5$, $+11\,55\,17$ &  $1230 \pm 1$ & $29 \pm 2$ & $0.77 \pm 0.04$ & $11.6$ & $7.63$ \\
Cloud 2  & $12\,13\,41.8$, $+12\,53\,51$ &  $2235 \pm 2$ & $53 \pm 3$  & $1.21 \pm 0.07$ & $9.2$  & $8.54$ \\

\enddata
\tablecomments{Col.~(1): cloud name. Col.~(2): right ascension and declination of cloud centroid (J2000). Col.~(3): average heliocentric velocity of integrated spectral profile from Figure~\ref{AOspectra}. Col.~(4): profile width, measured at 50\% of the integrated spectra profile peak and corrected for instrumental broadening as described in Giovanelli \etal (2007). Col.~(5): total flux of integrated spectral profile. Col.~(6): signal-to-noise ratio of the detection, computed using \wf\ and \fc\ via eq.~\ref{snr}. Col.~(7): base 10 logarithm of total HI mass, computed using \fc\ via eq.~\ref{HImass}.
}
\end{deluxetable}

\begin{deluxetable}{lcc}
\tablecaption{Aperture Synthesis Observing and Data Cube Parameters \label{VLAobs}}
\tablewidth{0pt}
\tabletypesize{\scriptsize}

\tablehead{ 
	\colhead{Parameter} & 
	\colhead{Cloud 1} &
	\colhead{Cloud 2}
}
\startdata
    Pointing center (J2000)                 & $12^{h}\,08^{m}\,45.5^{s}$, $+11^{\circ}\,55$\arcmin$\,17$\arcsec &  $12^{h}\,13^{m}\,41.8^{s}$, $+12^{\circ}\,53$\arcmin$\,51$\arcsec   \\
    Total time on-source                    &         547 minutes		&	532 minutes               	\\
    Net bandpass                            &  1.5~MHz (1132 -- 1336~\kms)    	&	1.5~MHz (2137 -- 2341~\kms) 	\\
    Maximum spectral resolution \chanp      &    12.2 kHz (2.6~\kms)       	&	12.2 kHz (2.6~\kms)		\\
    Synthesized beam / natural weighting    & $25.2'' \times 24.4''$ @ \,56.7\dg &	$25.3'' \times 24.4''$ @ \,57.2\dg\\
    $\smp$ at pointing center, $\chanp =2.6\,$\kms\      &      1.39 mJy/beam & 1.37 mJy/beam\\
\enddata
\end{deluxetable}

\newpage


\begin{deluxetable}{cccccccccc}
\tablecaption{Aperture Synthesis Cloud Properties from the VLA Observations \label{VLAparams}}
\tablewidth{0pt}
\tabletypesize{\scriptsize}
\tablehead{ 
	\colhead{Feature} & 
	\colhead{$(\alpha,\delta)^\prime$} & 
	\colhead{\czsunp } & 
	\colhead{$W_{50}^{\prime}$ } & 
	\colhead{\fcp } & 
	\colhead{\ahip} &
	\colhead{\pahip} &
	\colhead{$\log(\mhip/\msun)$} & 
	\colhead{$\log(\mdynp/\msun)$}  
	\\
	\colhead{} &
   \colhead{(J2000)} & 
   \colhead{(\kms)} & 
    \colhead{(\kms)} &  
    \colhead{(\jykms)} & 
    \colhead{(\arcmin)} &
    \colhead{(\arcdeg)} &
    \colhead{} & 
    \colhead{}
    \\
    \colhead{(1)} & 
    \colhead{(2)} &  
    \colhead{(3)} & 
    \colhead{(4)} &  
    \colhead{(5)} & 
    \colhead{(6)} & 
    \colhead{(7)} &
    \colhead{(8)} &
    \colhead{(9)}
}
\startdata
 Cloud 1 Global &  $-		$ 		 & $1229 \pm 2$ & $26 \pm 4$ & $0.67 \pm 0.03$ & $-          $  &    - & $7.64$ & $-   $ \\
 Cloud 1 North  &  $12\,08\,47.6$, $+11\,55\,57$ & $1234 \pm 3$ & $22 \pm 6$ & $0.29 \pm 0.03$ & $1.0 \pm 0.3$  & -268 & $7.28$ & $7.83$ \\
 Cloud 1 South  &  $12\,08\,47.4$, $+11\,54\,48$ & $1225 \pm 3$ & $20 \pm 8$ & $0.39 \pm 0.03$ & $1.4 \pm 0.3$  &   13 & $7.40$ & $7.65$ \\
 Cloud 2 Global &  $-           $ 		 & $2231 \pm 4$ & $51 \pm 6$ & $0.50 \pm 0.02$ & $-          $  &    - & $8.15$ & $-   $ \\
 Cloud 2 North  &  $12\,13\,42.5$, $+12\,54\,50$ & $2237 \pm 2$ & $13 \pm 4$ & $0.14 \pm 0.02$ & $2.5 \pm 0.04$ &   -7 & $7.60$ & $8.09$ \\
 Cloud 2 West   &  $12\,13\,33.1$, $+12\,52\,44$ & $2205 \pm 5$ & $41 \pm 9$ & $0.25 \pm 0.02$ & $2.4 \pm 0.03$ &   -4 & $7.85$ & $9.07$ \\
 Cloud 2 South  &  $12\,13\,41.9$, $+12\,51\,16$ & $2234 \pm 3$ & $ 6 \pm 5$ & $0.05 \pm 0.01$ & $0.8 \pm 0.02$ &   39 & $7.15$ & $7.00$ \\
\enddata
\tablecomments{Col.~(1): Cloud identifier. Col.~(2): centroid RA and Decl. based on the fitting of ellipses to each detection. Col.~(3): average heliocentric velocity of integrated spectral profile from Figure~\ref{VLAspectra}. Col.~(4): profile width, measured at 50\% of the integrated spectral profile peak and corrected for instrumental effects assuming that unbroadened profile is Gaussian. Col.~(5): total flux of integrated spectral profile. Col.~(6): maximum linear extent of region with $\nhp \geq 1.5 \times 10^{20}~\cm$ (Cloud 1 field) and $\nhp \geq 0.7 \times  10^{20}~\cm$ (Cloud 2 field) in the total intensity maps (Figs~\ref{C1moments}~and~\ref{C2moments}). Col.~(7): position angle at which \ahip\ was measured. Col.~(8): base 10 logarithm of total \hi\ mass, computed using \fcp\ via Equation~\ref{HImass}. Col.~(9): base 10 logarithm of the dynamical mass, computed using \wfp\ and \ahip\ via Equation~\ref{Dynmass}.
}
\end{deluxetable}

\begin{deluxetable}{lcrrlrll}
\tablecaption{Galaxy Environment of Cloud 1 ($cz_{\odot}$=1230~\kms) \label{Cloud1environ}}
\tablewidth{0pt}
\tabletypesize{\scriptsize}
\tablehead{ 
	\colhead{Galaxy Name} & 
	\colhead{$(\alpha,\delta)$} &
	\colhead{$cz_{\odot, HI}$ } & 
	\colhead{$cz_{\odot, optical}$} & 
	\colhead{NED Type} & 
	\colhead{$d_{C1}$} &
	\colhead{HI ref.} &
	\colhead{Optical ref.} 	
	\\
	\colhead{} &
   	\colhead{(J2000)} & 
   	\colhead{(\kms)} & 
   	\colhead{(\kms)} &  
   	\colhead{} & 
	\colhead{(kpc)} &
   	\colhead{} &
    	\colhead{} 
    	
    	\\
    	\colhead{(1)} & 
   	\colhead{(2)} &  
    	\colhead{(3)} & 
    	\colhead{(4)} &  
    	\colhead{(5)} & 
    	\colhead{(6)} & 
    	\colhead{(7)} &
	\colhead{(8)}
 
    	\\
}
\startdata
SDSS J120527.06+123243.2 &	120527.1,+123243 &		&	 771 &	Im	 &	297	&		&	AM06	\\
SDSS J120640.69+120204.3 &	120640.7,+120204 &		&       1428 &		 &	152	&		&	AM06	\\
AGC  226030		 &	120820.7,+123004 &	2320 	&	     &		 &	171	&	G07	&		\\
SDSS J120859.92+115631.2 &	120859.9,+115631 &		&	     &		 &	18	&		&		\\
AGC  224602		 &	121003.3,+114249 &	2557	&       2594 &		 &	111	&	K08	&	AM06	\\
VCC  20			 &	121018.8,+121949 &		&	     &		 &	163	&		&		\\
VCC  24			 &	121035.7,+114538 &	1296	&       1289 &	BCD	 &	139	&	K08	&	F99	\\
VCC  32			 &	121102.7,+120615 &		&       1894 &	E	 &	171	&		&	B85	\\
VCC  35			 &	121119.9,+115437 &		&	     &	 	 &	183	&		&		\\
SDSS J121146.77+122938.3 &	121146.8,+122938 &		&	 667 &		 &	272	&		&	AM06	\\	
VCC  41			 &	121204.4,+124408 &	2203	&	     &	IB	 &	335	&	B93	&		\\
VCC  46			 &	121210.9,+125335 &		&       1437 &	dE3	 &	374	&		&	AM06	\\
VCC  48			 &	121215.0,+122913 &         8	&        -53 &		 &	298	&	G07	&	B85	\\
VCC  58			 &	121232.3,+120723 &	2209	&       2213 &	SA(rs)bc &	276	&	G07	&	B85;R99	\\
IC   3041		 &	121242.7,+124546 &	1740	&       1738 &		 &	273	&	G07	&	AM06	\\	
VCC  65			 &	121243.2,+120719 &		&	     &	dE	 &	288	&		&		\\
\enddata
\tablecomments{
Col.~(1): Galaxy Name. 
Col.~(2): RA and Decl. reported in NED. 
Col.~(3): Heliocentric velocity based on HI measurements. 
Col.~(4): Heliocentric velocity based on optical spectroscopy. 
Col.~(5): NED morphological type. 
Col.~(6): Projected linear displacement from Cloud 1 at the Virgo distance of 16.7 Mpc.  
Col.~(7): HI measurement reference. 
Col.~(8): Optical measurement reference.  
References are abbreviated as:
G07: Giovanelli \etal 2007; K08: Kent \etal 2008; B85: Binggeli \etal 1985; B93: Binggeli \etal 1993; AM06: Adelman-McCarthy \etal 2006; F99: Falco \etal 1999 ; R99: Rubin \etal 1999
}
\end{deluxetable}

\begin{deluxetable}{lcrrlrll}
\tablecaption{Galaxy Environment of Cloud 2 ($cz_{\odot}$=2235~\kms)\label{Cloud2environ}}
\tablewidth{0pt}
\tabletypesize{\scriptsize}
\tablehead{ 
	\colhead{Galaxy Name} & 
	\colhead{$(\alpha,\delta)$} &
	\colhead{$cz_{\odot, HI}$ } & 
	\colhead{$cz_{\odot, optical}$} & 
	\colhead{NED Type} & 
	\colhead{$d_{C2}$} &
	\colhead{HI ref.} &
	\colhead{Optical ref.} 	
	\\
	\colhead{} &
   	\colhead{(J2000)} & 
   	\colhead{(\kms)} & 
   	\colhead{(\kms)} &  
   	\colhead{} & 
	\colhead{(kpc)} &
   	\colhead{} &
    	\colhead{} 
    	
    	\\
    	\colhead{(1)} & 
   	\colhead{(2)} &  
    	\colhead{(3)} & 
    	\colhead{(4)} &  
    	\colhead{(5)} & 
    	\colhead{(6)} & 
    	\colhead{(7)} &
	\colhead{(8)}
    	\\
}
\startdata
VCC 13 			&	120946.3,+133301&		&		&		&	337	&		&			\\
VCC 15			&	120954.5,+130258&	2535	&	2505	&       Sm	&	273	&	G07	&	F99		\\
VCC 20 			&	121018.8,+121949&		&		&		&	292	&		&			\\
VCC 22			&	121024.2,+131014&	1699	&	1726	&	BCD	&	247	&	G07	&	AM06		\\
VCC 23 			&	121025.3,+132155&		&		&		&	269	&		&			\\
AGC 224696		&	121038.0,+130119&	2394	&       2418	&		&	220	&	G07	&	AM06		\\
VCC 32			&	121102.7,+120615&		&	1894	&	E	&	298	&		&	B85		\\
VCC 35 			&	121119.9,+115437&		&		&		&	333	&		&			\\
VCC 36			&	121128.2,+133501&		&		&		&	255	&		&			\\
SDSS J121140.32+125824.6 &	121140.3,+125825&		&	2221	&		&	145	&		&	AM06		\\
SDSS J121141.87+131146.7 &	121141.9,+131147&		&	1496	&		&	166	&		&	AM06		\\
SDSS J121145.94+131707.9&	121145.9,+131708&		&	2468	&		&	178	&		&	AM06		\\
VCC 37			&	121146.2,+130124&		&	2308	&	dE5	&	142	&		&	AM06		\\
SDSS J121153.85+134830.2 &	121153.9,+134830&		&	2007	&		&	294	&		&	AM06		\\
VCC 41			&	121204.4,+124408&	2203	&		&	IB	&	125	&	B93	&			\\
VCC 46			&	121210.9,+125335&		&       1437	&	dE3	&	108	&		&	AM06		\\
VCC 47			&	121211.7,+131446&	1875	&	1890	&	SAB(r)a &	147	&	G07	&	AM06		\\
VCC 49			&	121217.2,+131218&		&	2295	&	E2	&	134	&		&	AM06		\\
VCC 55			&	121227.0,+131649&		&		&		&	142	&		&			\\
VCC 58			&	121232.3,+120723&	2209	&       2213	&	SA(rs)bc&	240	&	G07	&	B85		\\
IC 3041			&	121242.7,+124546&	1740	&       1738	&		&	 80	&	G07	&	AM06		\\
VCC 65			&	121243.2,+120719&		&		&		&	236	&		&			\\
VCC 68			&	121249.0,+132050&		&	2425	&		&	145	&		&	AM06		\\
VCC 70			&	121256.4,+130407&		&		&		&	73	&		&			\\
AGC 224807 		&	121309.4,+133504&	2100	&	2108	&		&	204	&	G07	&	AM06		\\
SDSS J121313.69+133122.0&	121313.7,+133122&		&	2158	&		&	185	&		&	AM06		\\
SDSS J121317.79+130935.6&	121317.8,+130936&		&	1915	&		&	82	&		&	AM06		\\
VCC 84			&	121335.3,+132413&		&		&		&	148	&		&			\\
VCC 85			&	121336.4,+130201&	1932    &		&		&	40	&	Ga06    &			\\
VCC 89			&	121347.3,+132528&	2114	&	2115	&	SAB(rs)cd &	154	&	G07	&	AM06		\\
VCC 97			&	121353.6,+131021&	2470	&	2480	&	SAB(s)c &	81	&	G07	&	F95		\\
VCC 98			&	121353.8,+135213&		&		&		&	284	&		&			\\
VCC 100			&	121404.0,+133908&		&		&		&	222	&		&			\\
VCC 106			&	121409.0,+115619&		&		&		&	281	&		&			\\
VCC 107			&	121410.7,+131407&		&		&		&	104	&		&			\\
SDSS J121419.86+132706.4& 	121419.9,+132706&		&	2467	&		&	168	&		&	AM06		\\
VCC 113			&	121432.8,+120611&	2115	&	2139	&		&	239	&	G07	&	AM06		\\
VCC 122			&	121444.2,+121048&		&	2348	&	S0	&	222	&		&	AM06		\\
AGC 224705		&	121444.6,+124723&	2279 	&	2298	&		&	81	&	G07	&	AM06		\\
VCC 132			&	121503.8,+130155&	2085	&	 	&	SB	&	105	&	G07	&			\\
VCC 133			&	121505.2,+130644&		&		&		&	117	&		&			\\
VCC 135			&	121506.8,+120058&	2402	&	2412	&	Sa	&	276	&	S05	&	AM06		\\
VCC 146			&	121520.8,+123656&		&		&		&	143	&		&			\\
VCC 150			&	121528.6,+123856&		&		&		&	146	&		&			\\
VCC 155   		&	121535.7,+133711&               &		&		&	250	&		&			\\
VCC 163			&	121546.0,+123344&		&		&		&	177	&		&			\\
VCC 164			&	121552.6,+120150&		&		&		&	296	&		&			\\
VCC 175			&	121602.8,+123544&		&		&		&	189	&		&			\\
VCC 185			&	121620.1,+130814&		&		&		&	200	&		&			\\	
VCC 197			&	121632.7,+130944&		&		&		&	216	&		&			\\
VCC 204			&	121639.2,+125220&	 	&		&		&	210	&		&			\\
VCC 215			&	121658.3,+121549&		&	2074	&	dE4	&	297	&		&	AM06		\\
VCC 224			&	121709.2,+122712&	2131	&	2109	&	Sbc	&	278	&	G07	&	AM06		\\
VCC 230			&	121719.5,+115632&		&	1429	&	dE4	&	380	&		&	AM06		\\
VCC 232			&	121723.6,+133020&		&		&		&	316	&		&			\\
AGC 224489		&	121728.1,+125556&	2056	&	2080	&		&	268	&	G07	&	AM06		\\
SDSS J121731.31+115715.9&       121731.3,+115716&   	        &		&		&	387	&		&			\\
VCC 241			&	121733.9,+122320&		&		&		&	312	&		&			\\

\enddata
\tablecomments{
Col.~(1): Galaxy Name. 
Col.~(2): RA and Decl. reported in NED. 
Col.~(3): Heliocentric velocity based on HI measurements. 
Col.~(4): Heliocentric velocity based on optical spectroscopy. 
Col.~(5): NED morphological type. 
Col.~(6): Projected linear displacement from Cloud 2 at the Virgo distance of 16.7 Mpc.  
Col.~(7): HI measurement reference. 
Col.~(8): Optical measurement reference.  
  References are abbreviated as:
G07: Giovanelli \etal 2007; 
K08: Kent \etal 2008; 
Ga06: Gavazzi \etal 2006;
B85: Binggeli \etal 1985; 
B93: Binggeli \etal 1993; 
AM06: Adelman-McCarthy \etal 2006; 
F99: Falco \etal 1999; 
R99: Rubin \etal 1999;
T08: Tully \etal 2008;
S05: Springob \etal 2005b;
F95: Fisher \etal 1995
}
\end{deluxetable}

\clearpage

\parskip 8pt

\onecolumn

\clearpage

\begin{thebibliography}{99}

{\scriptsize


\bibitem[Adelman-McCarthy et al.(2006)]{2006ApJS..162...38A} 
Adelman-McCarthy, J.~K., et al.\ 2006, \apjs, 162, 38 


\bibitem[Barnes et al.(2001)]{2001MNRAS.322..486B} Barnes, D.~G., et al.\ 
2001, \mnras, 322, 486 


\bibitem[Bell et al.(2003)]{2003ApJS..149..289B} Bell, E.~F., McIntosh, 
D.~H., Katz, N., \& Weinberg, M.~D.\ 2003, \apjs, 149, 289 


\bibitem[Binggeli et 
al.(1993)]{1993A&AS...98..275B} Binggeli, B., Popescu, C.~C., \& Tammann, G.~A.\ 1993, \aaps, 98, 275 


\bibitem[Binggeli et al.(1985)]{1985AJ.....90.1681B} Binggeli, B., Sandage, 
A., \& Tammann, G.~A.\ 1985, \aj, 90, 1681 


\bibitem[Cayatte et al.(1994)]{1994AJ....107.1003C} Cayatte, V., Kotanyi, 
C., Balkowski, C., \& van Gorkom, J.~H.\ 1994, \aj, 107, 1003 


\bibitem[Cayatte et al.(1990)]{1990AJ....100..604C} Cayatte, V., van 
Gorkom, J.~H., Balkowski, C., \& Kotanyi, C.\ 1990, \aj, 100, 604 


\bibitem[Chamaraux et 
al.(1980)]{1980A&A....83...38C} Chamaraux, P., Balkowski, C., \& Gerard, E.\ 1980, \aap, 83, 38 


\bibitem[Chung et al.(2007)]{2007ApJ...659L.115C} Chung, A., van Gorkom, 
J.~H., Kenney, J.~D.~P., \& Vollmer, B.\ 2007, \apjl, 659, L115 


\bibitem[Condon et al.(1998)]{1998AJ....115.1693C} Condon, J.~J., Cotton, 
W.~D., Greisen, E.~W., Yin, Q.~F., Perley, R.~A., Taylor, G.~B., 
\& Broderick, J.~J.\ 1998, \aj, 115, 1693 


\bibitem[Davies et al.(2004)]{2004MNRAS.349..922D} Davies, J., et al.\ 
2004, \mnras, 349, 922 


\bibitem[Doyon 
\& Joseph(1989)]{1989MNRAS.239..347D} Doyon, R., \& Joseph, R.~D.\ 1989, \mnras, 239, 347 


\bibitem[Falco et al.(1999)]{1999PASP..111..438F} Falco, E.~E., et al.\ 
1999, \pasp, 111, 438 


\bibitem[Fisher et al.(1995)]{1995ApJS..100...69F} Fisher, K.~B., Huchra, 
J.~P., Strauss, M.~A., Davis, M., Yahil, A., 
\& Schlegel, D.\ 1995, \apjs, 100, 69 


\bibitem[Ftaclas et al.(1984)]{1984ApJ...282...19F} Ftaclas, C., Struble, 
M.~F., \& Fanelli, M.~N.\ 1984, \apj, 282, 19 


\bibitem[Gavazzi et 
al.(2003)]{2003A&A...400..451G} Gavazzi, G., Boselli, A., Donati, A., Franzetti, P., \& Scodeggio, M.\ 2003, \aap, 400, 451 


\bibitem[Gavazzi et 
al.(2006)]{2006A&A...449..929G} Gavazzi, G., O'Neil, K., Boselli, A., \& van Driel, W.\ 2006, \aap, 449, 929 


\bibitem[Giovanelli 
\& Haynes(1985)]{1985ApJ...292..404G} Giovanelli, R., \& Haynes, M.~P.\ 1985, \apj, 292, 404 


\bibitem[Giovanelli 
\& Haynes(1983)]{1983AJ.....88..881G} Giovanelli, R., \& Haynes, M.~P.\ 1983, \aj, 88, 881 


\bibitem[Giovanelli 
\& Haynes(1989)]{1989ApJ...346L...5G} Giovanelli, R., \& Haynes, M.~P.\ 1989, \apjl, 346, L5 


\bibitem[Giovanelli et al.(2005)]{2005AJ....130.2613G} Giovanelli, R., et 
al.\ 2005a, \aj, 130, 2613 


\bibitem[Giovanelli et al.(2005)]{2005AJ....130.2598G} Giovanelli, R., et 
al.\ 2005b, \aj, 130, 2598 


\bibitem[Giovanelli et al.(2007)]{2007AJ....133.2569G} Giovanelli, R., et 
al.\ 2007, \aj, 133, 2569 


\bibitem[Greisen(2003)]{2003ASSL..285..109G} Greisen, E.~W.\ 2003, 
Astrophysics and Space Science Library, 285, 109 


\bibitem[Haynes et al.(2007)]{2007ApJ...665L..19H} Haynes, M.~P., 
Giovanelli, R., \& Kent, B.~R.\ 2007, \apjl, 665, L19 


\bibitem[Helou et al.(1984)]{1984ApJS...55..433H} Helou, G., Hoffman, 
G.~L., \& Salpeter, E.~E.\ 1984, \apjs, 55, 433 


\bibitem[Henning et 
al.(1993)]{1993A&A...268..536H} Henning, P.~A., Sancisi, R., \& McNamara, B.~R.\ 1993, \aap, 268, 536 


\bibitem[Kent et al.(2008)]{2008AJ....136..713K} Kent, B.~R., et al.\ 2008, 
\aj, 136, 713 


\bibitem[Kent et al.(2007)]{2007ApJ...665L..15K} Kent, B.~R., et al.\ 2007, 
\apjl, 665, L15 


\bibitem[Kent et al.(2009)]{2009ApJ...691.1595K} Kent, B.~R., Spekkens, K., 
Giovanelli, R., Haynes, M.~P., Momjian, E., Cort{\'e}s, J.~R., Hardy, E., 
\& West, A.~A.\ 2009, \apj, 691, 1595 


\bibitem[Kent(2008)]{2008PhDT.........5K} Kent, B.~R.\ 2008, Ph.D.~Thesis, Cornell University


\bibitem[Kilborn et al.(2000)]{2000AJ....120.1342K} Kilborn, V.~A., et al.\ 
2000, \aj, 120, 1342 


\bibitem[Klypin et al.(1999)]{1999ApJ...522...82K} Klypin, A., Kravtsov, 
A.~V., Valenzuela, O., \& Prada, F.\ 1999, \apj, 522, 82 


\bibitem[Kniazev et al.(2004)]{2004AJ....127..704K} Kniazev, A.~Y., Grebel, 
E.~K., Pustilnik, S.~A., Pramskij, A.~G., Kniazeva, T.~F., Prada, F., 
\& Harbeck, D.\ 2004, \aj, 127, 704 


\bibitem[Koopmann et al.(2008)]{2008ApJ...682L..85K} Koopmann, R.~A., et 
al.\ 2008, \apjl, 682, L85 


\bibitem[Masters et al.(2004)]{2004ApJ...607L.115M} Masters, K.~L., Haynes, 
M.~P., \& Giovanelli, R.\ 2004, \apjl, 607, L115 


\bibitem[McLaughlin(1999)]{1999ApJ...512L...9M} McLaughlin, D.~E.\ 1999, 
\apjl, 512, L9 


\bibitem[Meyer et al.(2004)]{2004MNRAS.350.1195M} Meyer, M.~J., et al.\ 
2004, \mnras, 350, 1195 


\bibitem[Minchin et al.(2005)]{2005ApJ...622L..21M} Minchin, R., et al.\ 
2005, \apjl, 622, L21 


\bibitem[Moore et al.(1999)]{1999ApJ...524L..19M} Moore, B., Ghigna, S., 
Governato, F., Lake, G., Quinn, T., Stadel, J., 
\& Tozzi, P.\ 1999, \apjl, 524, L19 


\bibitem[Oosterloo 
\& van Gorkom(2005)]{2005A&A...437L..19O} Oosterloo, T., \& van Gorkom, J.\ 2005, \aap, 437, L19 


\bibitem[Roberts(1988)]{1988nda..conf...65R} Roberts, M.~S.\ 1988, New 
Ideas in Astronomy, 65 


\bibitem[Roberts(1975)]{1975gaun.book..309R} Roberts, M.~S.\ 1975, Galaxies 
and the Universe, 309 


\bibitem[Roberts et al.(2007)]{2007MNRAS.379.1053R} Roberts, S., Davies, 
J., Sabatini, S., Auld, R., \& Smith, R.\ 2007, \mnras, 379, 1053 


\bibitem[Rosenberg 
\& Schneider(2002)]{2002ApJ...567..247R} Rosenberg, J.~L., \& Schneider, S.~E.\ 2002, \apj, 567, 247 


\bibitem[Rubin et al.(1999)]{1999AJ....118..236R} Rubin, V.~C., Waterman, 
A.~H., \& Kenney, J.~D.~P.\ 1999, \aj, 118, 236 


\bibitem[Ryder et al.(2001)]{2001ApJ...555..232R} Ryder, S.~D., et al.\ 
2001, \apj, 555, 232 


\bibitem[Salzer et al.(1991)]{1991AJ....101.1258S} Salzer, J.~J., di Serego 
Alighieri, S., Matteucci, F., Giovanelli, R., 
\& Haynes, M.~P.\ 1991, \aj, 101, 1258 


\bibitem[Sancisi et al.(1987)]{1987ApJ...315L..39S} Sancisi, R., Thonnard, 
N., \& Ekers, R.~D.\ 1987, \apjl, 315, L39 


\bibitem[Schneider et al.(1983)]{1983ApJ...273L...1S} Schneider, S.~E., 
Helou, G., Salpeter, E.~E., \& Terzian, Y.\ 1983, \apjl, 273, L1 


\bibitem[Shibata et al.(2001)]{2001ApJ...549..228S} Shibata, R., 
Matsushita, K., Yamasaki, N.~Y., Ohashi, T., Ishida, M., Kikuchi, K., 
B{\"o}hringer, H., \& Matsumoto, H.\ 2001, \apj, 549, 228 


\bibitem[Snowden et al.(1995)]{1995ApJ...454..643S} Snowden, S.~L., et al.\ 
1995, \apj, 454, 643 


\bibitem[Solanes et al.(2001)]{2001ApJ...548...97S} Solanes, J.~M., 
Manrique, A., Garc{\'{\i}}a-G{\'o}mez, C., Gonz{\'a}lez-Casado, G., 
Giovanelli, R., \& Haynes, M.~P.\ 2001, \apj, 548, 97 


\bibitem[Springob et al.(2005)]{2005ApJ...621..215S} Springob, C.~M., 
Haynes, M.~P., \& Giovanelli, R.\ 2005, \apj, 621, 215 


\bibitem[Springob et al.(2005)]{2005ApJS..160..149S} Springob, C.~M., 
Haynes, M.~P., Giovanelli, R., \& Kent, B.~R.\ 2005, \apjs, 160, 149 


\bibitem[Tonry et al.(2000)]{2000ApJ...530..625T} Tonry, J.~L., Blakeslee, 
J.~P., Ajhar, E.~A., \& Dressler, A.\ 2000, \apj, 530, 625 


\bibitem[Tully et al.(2008)]{2008ApJ...676..184T} Tully, R.~B., Shaya, 
E.~J., Karachentsev, I.~D., Courtois, H.~M., Kocevski, D.~D., Rizzi, L., 
\& Peel, A.\ 2008, \apj, 676, 184 


\bibitem[Vollmer et al.(2001)]{2001ApJ...561..708V} Vollmer, B., Cayatte, 
V., Balkowski, C., \& Duschl, W.~J.\ 2001, \apj, 561, 708 


\bibitem[Wong et al.(2006)]{2006MNRAS.371.1855W} Wong, O.~I., et al.\ 2006, 
\mnras, 371, 1855 


\bibitem[York et al.(2000)]{2000AJ....120.1579Y} York, D.~G., et al.\ 2000, 
\aj, 120, 1579 





}

\end{thebibliography}
\end{document}